\documentclass[%
 aip,
 amsmath,amssymb,
 reprint,%
]{revtex4-1}

\usepackage{graphicx}
\usepackage{dcolumn}
\usepackage{bm}

\usepackage[utf8]{inputenc}
\usepackage[T1]{fontenc}
\usepackage{mathptmx}
\usepackage{etoolbox}

\makeatletter
\def\@email#1#2{%
 \endgroup
 \patchcmd{\titleblock@produce}
  {\frontmatter@RRAPformat}
  {\frontmatter@RRAPformat{\produce@RRAP{*#1\href{mailto:#2}{#2}}}\frontmatter@RRAPformat}
  {}{}
}%
\makeatother
\begin{document}


\title{Feasibility of the Josephson voltage and current standards on a single chip}

\author{Rais S. Shaikhaidarov}
\email{R.Shaikhaidarov@rhul.ac.uk}
\affiliation{Royal Holloway, University London, Egham, TW20 0EX, Surrey UK} 
\affiliation{National Physical Laboratory, Hampton Road, Teddington, TW11 0LW,  UK}

\author{Ilya Antonov}%
\affiliation{Royal Holloway, University London, Egham, TW20 0EX, Surrey UK}%

\author{Kyung Ho Kim}
\affiliation{Royal Holloway, University London, Egham, TW20 0EX, Surrey UK}%

\author{Artem Shesterikov}
\affiliation{Royal Holloway, University London, Egham, TW20 0EX, Surrey UK}
\affiliation{National Physical Laboratory, Hampton Road, Teddington, TW11 0LW,  UK}

\author{Sven Linzen}
\affiliation{Leibniz Institute of Photonic Technology, D-07702 Jena, Germany}

\author{Evgeni V. Il’ichev}
\affiliation{Leibniz Institute of Photonic Technology, D-07702 Jena, Germany}

\author{Vladimir N Antonov}
\affiliation{Royal Holloway, University London, Egham, TW20 0EX, Surrey UK}%

\author{Oleg V Astafiev}
\affiliation{Royal Holloway, University London, Egham, TW20 0EX, Surrey UK}
\affiliation{Skolkovo Institute of Science and Technology, Bolshoy Boulevard 30, Moscow, Russia}

\date{\today}

\begin{abstract}
The quantum Josephson voltage standard is well established across the metrology community for many years. It relies on the synchronisation of the flux tunneling in the S/I/S Josepson junctions (JJ) with the microwave radiation (MW) of frequency $f$ such that $V=\Phi_0 fm$ where $m$=~0, 1, 2, $\cdots$. The phenomenon is called the Shapiro steps. Together with the Quantum Hall resistance standard, the voltage standard forms the foundation of electrostatic metrology. The current is then defined as the ratio of the voltage and resistance. Realisation of the quantum current standard, would close the electrostatic metrological triangle of voltage-resistance-current. The current quantisation $I=2efm$, the inverse Shapiro steps, was recently shown using the superconducting nanowires and small JJ. The effect is a synchronization of the MW with the Cooper pair tunnelling. This paves the way to combine the JJ voltage and current standards on the same chip and demonstrate feasibility of the multi-standard operation. We show the voltage and current quantization on the same chip up to frequency of 10 GHz, corresponding to the amplitudes $\sim$20.67~$\mu$V and $\sim$3.23 nA respectively. The accuracy of the voltage and current quantisation, however, is relatively low, 35 ppk and 100 ppk respectively. We discuss measures to optimise the JJs, circuit and environment to boost the amplitude and accuracy of the standards.  
\end{abstract}

\maketitle

%


The electricity has a metrological triangle: the voltage, resistance and current. Two of them, the voltage and resistance, are defined with high accuracy, down to a few parts per billion. The current is determined in practice as the ratio of standard voltage to standard resistance. However, there is a need to close the metrological triangle with a quantum current standard with the same accuracy as the first two. The current standard can be realised using the periodic current transfer device $I=efm$, with the frequency of charge transfer $f$ and $m$=~0, 1, 2, $\cdots$. There are a few types of devices that can meet the current standard. A charge transfer through the semiconductor quantun dot (QD) and single electron transistor (SET) turnstile with the drive frequency $f$ are among them \cite{Giblin2020,Pekola2013,Crescini2023}. The highest accuracy of the current standard demonstrated so far is done by the QD device. The charge can be transferred with an accuracy of about one ppm with a driving frequency of a few GHz. A further increase in the driving frequency and accuracy of the current state is limited by non-adiabatic excitations of the electrons in the QD. Recently another realization of the quantum current standard was demonstrated, where the current quantization in the superconducting nanowires $I_m=2efm$ was observed \cite{Shaikh2022}. The effect is a consequence of the Coherent Quantum Phase Slip (CQPS) in the superconductor \cite{Mooij06}.  The accuracy of quantization, however, conceded to that of the QD device, being only few ppk. But the top frequency demonstrated with CQPS device is extended to 30 GHz. Further development of the device and accuracy is believed to be feasible. The fabrication technology of CQPS device is compatible to that of the voltage standard based on the JJs so that it is possible to combine two standards on the same chip. The hurdle is, however, the yield of the devices. Fabrication of the superconducting nanowires with the proper CQPS parameters is a challenging task. The superconducting nano-wires of 20-30 nm long and $\sim$10 nm wide is the art of the modern nanofabrication. Even to be geometrically identical the wires of this size have a wide variation of the superconducting parameters, critical current and CQPS energy. Recent breakthrough in demonstration of the current quantization in small JJ has eased the problem~\cite{Shaikh2024,kaap2024demonstration}. Technology of JJ fabrication is well established and ensures the higher yield. We combine two standards, the voltage and current, on one chip and demonstrate their operation with one source of the MW. In this work we present experimental data and discuss further development of the double-standard chip.
    
\begin{figure}
    \includegraphics[scale=0.37]{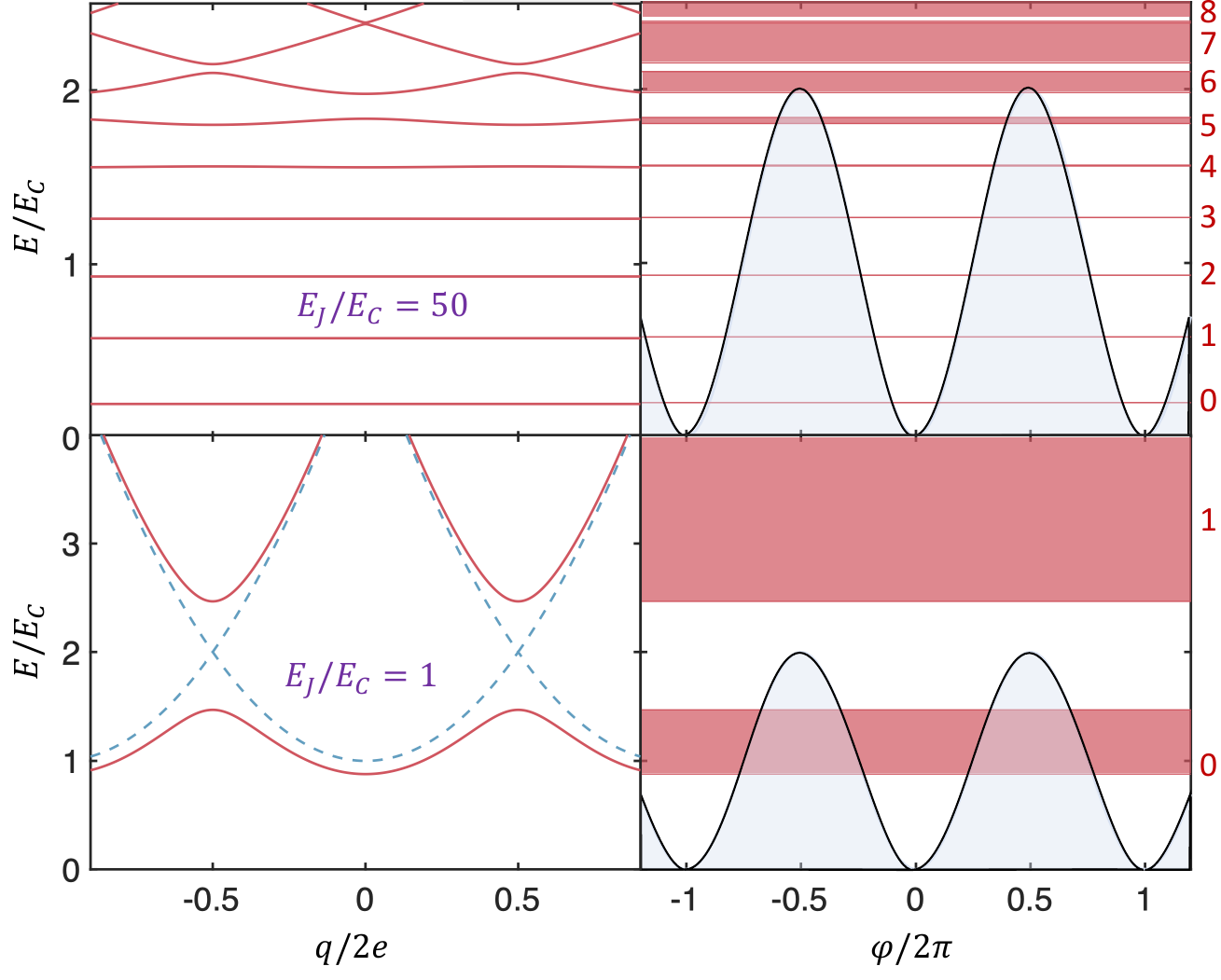}
    \caption{The Bloch bands of the JJ in two regimes: $E_J>E_C$ (top) and $E_J<E_C$ (bottom). The regime at the top corresponds to the direct Shapiro steps, $V=(h/2e)fm$, while at the bottom corresponds to the inverse Shapiro steps, $I=2efm$. The red numbers on the right correspond to the energy levels/bands of the system. The blue dashed lines are charge states with $E_J = 0$.}
    \label{fig:shapiro1}
\end{figure}

There are few types of the JJs useful for different applications. They can be distinguished by the state of the conjugated variables of the JJ, the charge $q$ at the junction capacitor and the superconducting phase $\varphi$ across the junction. When the Josephson energy of the JJs, $E_J=I_c\Phi_0/\pi$, is much larger than the charging energy, $E_C=e^2/2C$ ($I_C$ and $C$ are the critical current and capacitance of the JJ), the phase $\phi$ is well defined. The voltage steps, Shapiro steps (SS), $V_m=\Phi_0fm$ ($m$=1, 2..., $\Phi_0=h/2e$) are developed under the MW of frequency $f$ at the $I-V$ curve  \cite{Shapiro1963}. The effect is used for the voltage standard of high accuracy, $\sim$0.1 ppb. In the opposite limit $E_C~>>~E_J$ the charge $q$ at the JJ is well defined, and the device operates as the Single Electron Transistor, the sensitive electrometer and potentiometer \cite{Kleinschmidt06}. There is one more useful type of the JJ: it was theoretically predicted that when $E_C~\sim~E_J$ the current steps, the inverse Shapiro steps (ISS), $I_m=2efm$ should be developed at the voltage biased $I-V$ curve under the MWs \cite{Averin1985}. The effect has been recently demonstrated experimentally ~\cite{Shaikh2024,kaap2024demonstration}. It can be used for development of the quantum standard of current. Energy diagram in Fig.~\ref{fig:shapiro1} explains the underlying physics of the two types of the JJ leading to the voltage and current standards. At the top diagram the energy of the system normalised to $E_J$ is calculated in coordinates of the charge $q/2e$ and phase $\phi/2\pi$ when $E_J/E_C$=50. The modulation of the energy with charge forms the Bloch bands (Bloch oscillations), numbered with 0, 1, 2,.... They are shown as the colour stripes at the right panel. The lowest Bloch bands, 0 and 1, is well below the potential barriers of the system in the right panel, ensuring the well defined phase. MW photon assisted tunneling of the fluxes thorough these barriers gives rise to the quantized voltage across the JJ, the direct Shapiro steps. In the case $E_J/E_C$=1, the lower part in the figure, one has the coherent flux tunnelling, but the charge transfer is suppressed by the barriers, seen in the left panel. The height of these barriers is the CQPS energy $E_S$:
\begin{eqnarray}
E_{S} =\sqrt{\frac{8\eta}{\pi}}E_p e^{-\eta},
\label{ES}
\end{eqnarray}
where $E_p=\sqrt{8E_JE_C}$ is the plasma energy of the JJ and $\eta=E_p/E_C=\sqrt{8E_J/E_C}$. The charge tunneling is promoted when the Bloch oscillations are synchronized with the MW. Then the quantized current steps, or the inverse Shapiro steps, are developed at the $I-V$ curve.

The differential equations describing evolution of the phase and charge in two cases are:
\begin{eqnarray}
CR\ddot\varphi + \dot\varphi + \frac{I_C}{2e}\sin{2\pi\varphi}= \frac{R_NI_{\rm dc}}{\Phi_0} + \frac{R_NI_{\rm ac}}{\Phi_0}\cos(2\pi ft)
\label{eq:direcShapiro}\\
\frac{L}{R_N}\ddot q + \dot q + \frac{V_C}{R_N}\sin\frac{2\pi q}{2e} = \frac{V_{\rm dc}}{R_N} + \frac{V_{\rm ac}}{R_N}\cos(2\pi ft)+\xi(t)
\label{eq:inverseShapiro}
\end{eqnarray}
In these equations $R_N$ is the normal resistance of the JJ, $V_C = \pi E_S/e$ is the critical voltage for the CQPS, $\xi(t)$ is the noise current, and $I_{\rm ac}$ and $V_{\rm ac}$ are the amplitudes of the MW. The noise current is important for (\ref{eq:inverseShapiro}) as it is comparable with $V_c/R_N$ in our experiment. Equations (\ref{eq:direcShapiro}) and (\ref{eq:inverseShapiro}) have a stationary dual solutions of the quantized current and voltage steps 
\begin{eqnarray}
I(V_{\rm dc},V_{\rm ac})=\sum_m{J_m\left(\frac{V_{\rm ac}}{\Phi_0 f}\right) I_0(V_{\rm dc}-\Phi_0 fm)}
\label{eq:directsteps}\\
V(I_{\rm dc},I_{\rm ac})=\sum_m{J_m^2\left(\frac{I_{\rm ac}}{2ef}\right) V_0(I_{\rm dc}-2efm)}
\label{eq:inverssteps}
\end{eqnarray}
One can notice that that the current steps are modulated with the square of Bessel function in (\ref{eq:inverssteps}) compared to the plain Bessel function in SS. The effect is a consequence of accounting of the thermal noise as we are working in the regime of $E_S<k_BT$ \cite{Shaikh2024}.    
It is intriguing to combine the circuits for the two effects, the SS and ISS,  on one chip by designing JJs of different parameters. This would simplify the metrological calibration and may offer a new functionality for the quantum technology platform. 

\begin{figure}
    \includegraphics[scale=0.8]{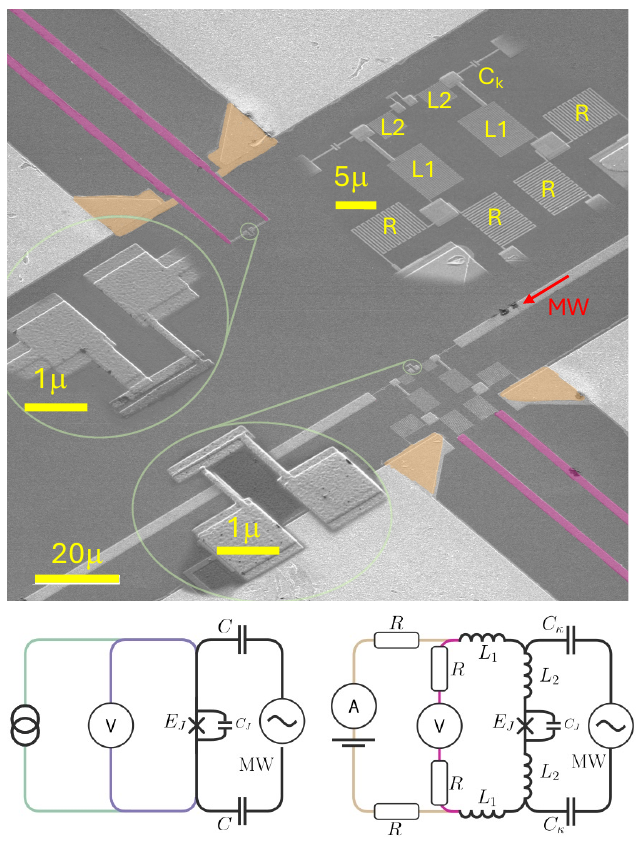}
    \caption{Focused Ion Beam photo of the chip for SS (top ) and ISS (bottom). The colour leads are $V+/-$ and $I+/-$ for the four-point measurement. The JJs of the SS and ISS have different sizes (inserts at the left): 700~x~100~nm$^2$ and 80~x~40~nm$^2$. The ISS circuit ( insert at the right) has Pd resistors $R$= 6.3~k$\Omega$ and TiN superinductances $L1$+$L2$= 1.49 ~$\mu$H to screen the JJ from the environmental noise. The MW is supplied by the lead through the capacitor $C_k\approx$~0.1~fF. The main picture and inserts have their own scale bars. The measurement circuits for the SS (left) are ISS (right) are shown at the bottom.}
    \label{fig:photo}
\end{figure}

In this work we demonstrate feasibility of the chip combining the voltage and current standards. The top and bottom circuit in Fig.~\ref{fig:photo} are designed for the the SS and ISS respectively. The JJs in both circuits are made of Al using a shadow deposition technique. The junctions have different size, 700$\times$100~nm$^2$ and 80$\times$40~nm$^2$. Different parameters of Al oxidation are also used when an insulating layer is formed in the JJs. We have measured four samples and present data for one of them with $E_J/E_C\sim$4.8 ($E_J$=~26.9~GHz, $E_C$=~5.5~GHz) for the SS and $E_J/E_C\sim$0.7 ($E_J$=~84.5~GHz, $E_C$=~121~GHz)for the ISS. The sample is placed in a copper box at 15~mK stage of the dilution refrigerator. Four $dc$ leads, $I+, I-, V+, V-$, connect JJs to the four-probe measurement scheme. The leads are thermalised and filtered at the different stages of the refrigerator, on the way from the room temperature to 15~mK stage. Finally, the leads are passed through a box with a cascaded Low Temperature Co-Fired Ceramic (LTCC) low-pass filters with a stop band from 80MHz to 20~GHz. The ISS  has a further isolation from the environment with the Pd resistors $R$=~6.3~k$\Omega$ and TiN superinductances $L1+L2$=~1.15+0.34~$\mu$H. There are traps next to the $L1$ and $L2$ to eliminate non-equilibrium qusi-particles generated by the MW. All these measures are taken for the ISS because the CQPS energy $E_S$ is smaller than the thermal energy $k_BT$, promoting the thermal current noise $\xi(t)$ in (\ref{eq:inverseShapiro}). The issue of the noise and $E_S$ is discussed in \cite{Shaikh2024}. 

\begin{figure}
    \includegraphics[scale=0.7]{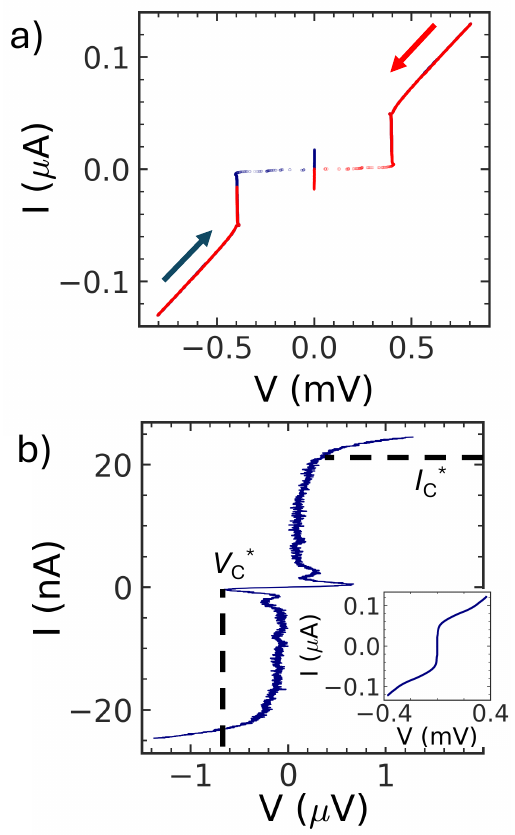}
    \caption{$I-V$ curves of the SS and ISS. a) The SS is in under-damped regime with $I_C$=~50~nA b) The ISS has the current blockade below $V_C^*$=~0.65$\mu$V and apparent critical current $I_C^*$ =~21~nA. Wide range $I-V$ curve is shown in the insert.}
    \label{fig:IV}
\end{figure}

The normal resistance for the SS and ISS JJs are 5.9~k$\Omega$ and 1.9~k$\Omega$ (41.3~k$\Omega$ and 0.6~k$\Omega$ per area of 100$\times$100~nm$^2$). Different resistance per square reflects different oxidation parameters during fabrication of the JJs. The $I-V$ curves of the SS and ISS are shown in Fig.~\ref{fig:IV}. The curves resembles each other at the large scale, but they are quite different at low voltage/current bias: while there is a pure superconducting branch at SS curve with $I_C\approx$54~nA, Fig.~\ref{fig:IV}(a), the ISS has the current blockade below 0.65~$\mu$V followed by the recovery to the state with resistance close to zero, Fig.~\ref{fig:IV}(b). The apparent critical current of the ISS, $I_C^*$, shown in the figure, is around 21~nA. It is below 15$\%$ of the value expected from the Ambegoakar-Baratov equation,  $I_C=\pi\Delta/2eR_N\sim$170~nA, when taking the superconducting gap of Al as 200~$\mu$V. We believe that the critical current is suppressed in the ISS due to Zener tunneling \cite{Shaikh2024}. 
\begin{eqnarray}
I_Z=\frac{\pi E_J}{16E_{C}}I_C
\label{Zener}
\end{eqnarray}
One gets $I_Z$=~23~nA when parameters of the ISS are plugged into (\ref{Zener}), which is fairly close to the experimental value. The  McCumber parameters $\beta$ of the JJs are 20 and 0.29 respectively. The McCumber parameter is used to characterize the JJ suitable for the voltage standard. To have accurate voltage steps, the $\beta$ should be much larger than one, which is fully satisfied in our experiment. The relevant parameter for the observation of the ISS is $E_J/E_C$, which should be around one. The ISS shown here satisfies this criteria, $E_J/E_C$=~0.7. One can find a discussion of the optimisation of the ISS parameters in a recent publication \cite{Shaikh2024}. 

\begin{figure}
    \includegraphics[scale=0.55]{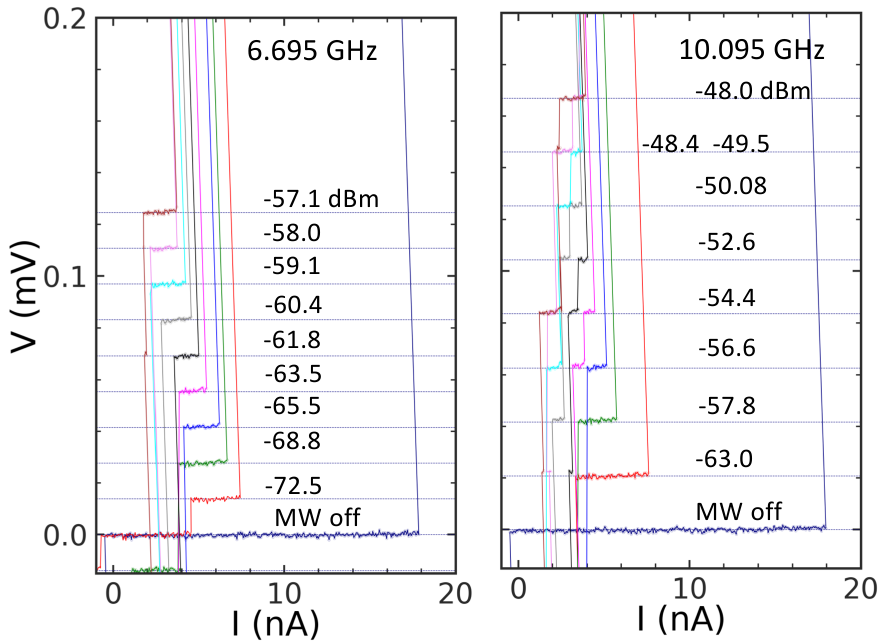}
    \caption{Direct Shapiro steps at 6.695~GHz and 10.095~GHz. The curves are taken at different MW amplitude so that the voltage steps of different orders $m$ are present.  The horizontal lines are the guide lines for the expected position of the voltage steps $V_m=\Phi_0fm$. The most right curves are taken without the MW.}
    \label{fig:expShap}
\end{figure}

The MW is applied to the devices by the co-planar waveguide terminating with the 10 $\mu$m wide lead capacitively coupled to the ISS through the $C_k\approx$0.1~fF, see the right insert in Fig.~\ref{fig:photo}. There is no special arrangement to couple the MW to the SS, so the same MW line is used to drive the SS. The microwave is coupled to SS by the radiated electric field of the microwave line. The voltage steps are easily produced  in the SS under the MW, Fig.~\ref{fig:expShap}. The horizontal lines in the figure indicate expected positions of the voltage steps. The curves with different steps, $m$ changes from 0 to 10, are taken at different amplitude of the MW. The width of the voltage steps is modulated by the MW amplitude (\ref{eq:directsteps}). Accuracy of the voltage steps, a dimensionless deviation of the voltage from $\Phi_0fm$, is around 35~ppk. There are a few reasons for the low accuracy of the voltage steps. One of them is the non-optimal $dc$ circuit used in the experiment. The measurements should be done using a pure current bias scheme. We use a universal amplifier, where one can change from the voltage to current bias by changing of the bias resistors, $R_b$ \cite{Shaikh2022}. The pure current bias circuit can not be realised with this setup. In the experiment we use a mixed bias measurements with $R_b$=100~k$\Omega$. The current and voltage across the sample are varied with the bias and both are recorded during the experiment. 

\begin{figure}[h!]
    \includegraphics[scale=0.65]{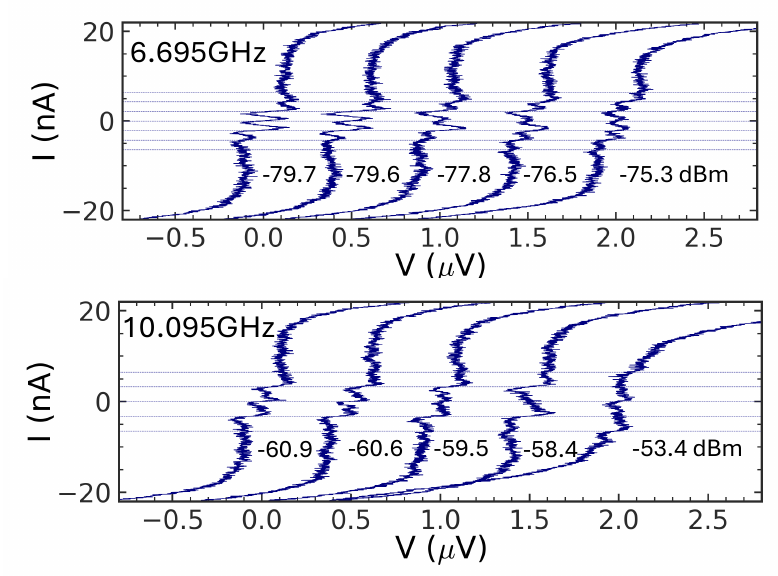}
    \caption{Inverse Shapiro steps at 6.695~GHz and 10.095GHz. The curves are taken at different MW amplitude. The horizontal lines are the guide lines for the expected position of the current steps $I_m=2efm$.}
    \label{fig:expinvShap}
\end{figure}

The ISS circuit develops the current steps when the MW is applied, Fig.~\ref{fig:expinvShap}. Similar to the SS the steps of higher order appear when a stronger MW drive is applied. The steps are modulated with the MW amplitude as square of the Bessel function. The accuracy of the current steps is only 100~ppk. In the next section, we discuss possible measures to improve accuracy.

 In our experiments the voltage and current step have accuracy well below the metrological needs. The way to improve SS, is straightforward. Currently the voltage standard uses the Nb technology instead of Al one. Nb has the superconducting gap, $\delta$, of 1.4~meV, which is about seven time larger than that of Al. As a result, Nb SS has larger critical current, and it is less acceptable to the environmental noise: the arbitrary flux tunnelling across the JJ is strongly suppressed. The circuitry, electronics and screening environment of the voltage standard is also well elaborated, which can be invoked for the metrological chip.

 ISS operation and its technology are less developed. They are currently under an intensive development. One of the issues is the relatively low value of the CQPS energy $E_S$ compared to the temperature of the electron bath. We can estimate the energy from the apparent critical voltage as $E_S\sim eV_C^*/\pi h$=~50~MHz, which corresponds to temperature of 3 mK. It is smaller than the base temperature of the experiment. The quantized current plateaus are still present, but their accuracy concedes substantially to that of SS. The estimation of the $E_S$ from (\ref{ES}) using parameters of the JJ gives 65.7~GHz, which is three orders of magnitude higher. The nature of such a strong suppression is not fully understood at the moment. It can be attributed to the leak of electro-magnetic noise to the JJ from the environment \cite{Shaikh2022}. The resonances at the $I-V$ curve without MW illumination in Fig.~\ref{fig:IV}(b) indicate at this. Re-designing geometry and parameters of the screening circuit may suppress unwanted resonances. Another factor to consider is the noise current, $\delta I_t$ produced by the screening resistors $R$ heated by the MW. The estimation of $\delta I_t$ gives $\sim$1~nA \cite{Shaikh2022}, which is comparable with the quantized current itself. It is possible to design a circuit that delivers the MW more effectively to the JJ, reducing the MW power used in the experiment. Then the overall value of the MW power used in the experiment can be reduced. A concept with more effective coupling of the MW to JJ is recently suggested by Erdmanis and Nazarov \cite{Erdmans2022}. They suggest to split the JJ to two, with a small metal island in between, so that the current quantization can be controlled with the gate electrode capacitively coupled to the island. The same gate electrode can be used to deliver the MW in a highly efficient way. One can also cool down the chip together with the resistors $R$ in the $^3$He bath, so called immersion cooling \cite{Lucas2023}. This may reduce temperature of $R$ by factor of $(\delta T/T)^{1/5}$ and, consequently, the amplitude of the noise current $\delta I_t$. Despite the cooling factor is small, it may have a substantial effect on accuracy of the current quantization because of an exponential sensitivity of the effect to ratio of $E_S$ to the noise power.

There is a room also in the material research. The prime candidate material for the ISS is Nb. Nb has higher $E_J$, and, correspondingly $E_S$, because of a larger superconducting gap. This, however, should be done in pair with reducing the size of the JJ to keep ratio $E_J/E_C$ close to one. The latter may encounter difficulty in nanofabrication, as the Nb technology is less prone for small JJ.

Finally, one must utilise a pure voltage bias scheme for the ISS. Currently there are two symmetric 100~ k$\Omega$ bias resistors in the differential amplifier, which somewhat limit the current and have an adverse effect on the flatness of the plateau (ISS resistance at the plateau when using the mixed bias scheme is $\sim$~3~k$\Omega$. One would expect substantial improvement of the quantization with the pure voltage bias scheme.

We report on demonstration of the direct and inverse Shapiro steps on one chip. The research indicates at feasibility of development of the metrological chip having both, the voltage and current standard. There is, however, issue with the accuracy of the quantization, particular for the current standard. We suggest few directions to tackle the problem. The research in these directions is currently under way.

This work was supported by Engineering and Physical Sciences Research Council
(EPSRC) Grant No. EP/T004088/1, European Union's Horizon 2020 Research and
Innovation Programme under Grant Agreement No. 862660/Quantum E-Leaps and
20FUN07 SuperQuant.
\bibliography{Two_standards_APL}
\end{document}